\documentclass{appolb}
\usepackage{graphicx}
\usepackage{subfigure}
\usepackage{amsmath}
\pdfoutput=1

\begin{document}
\title{Populating Low-Spin States in Radioactive Nuclei to Measure Magnetic Moments Using the Transient Field Technique%
\thanks{Presented at Jagiellonian Symposium on Fundamental and Applied Subatomic Physics, Krakow, Polland, 2015.}%
}
\author{D.A. Torres, F. Ram\'{\i}rez
\address{Departamento de F\'{\i}sica, Universidad Nacional de Colombia, Bogot\'a, Colombia}
\\
}
\maketitle
\begin{abstract}
The experimental study of magnetic moments  for nuclear states
near the ground state, $I \ge 2$, provides a powerful tool to test nuclear
structure models. The study of magnetic moments in nuclei far away from the stability line is the next frontier in such studies. Two techniques have been utilized to populated low-spin states in radioactive nuclei: coulomb excitation reactions using radioactive nuclei, and the transfer of $\alpha$ particles to stable beams to populate low spin states in radioactive nuclei. A presentations of these two techniques, along with the experimental challenges presented for future uses with nuclei far away from the stability line, will be presented.
\end{abstract}

\section{Introduction}
The arise of new radioactive beam facilities around the world has open the possibility to explore the limits of the nuclear matter well beyond the stability line. The characterization of the neutron and proton components of the wave function of such radioactive nuclei has a very important tool with the measurements of magnetic moments. The so called Transient Field Technique (TF)~\cite{0954-3899-34-9-R01,kumTF2015} provides integral information of the magnetic moment, with a value and a sign for the nuclear $g$ factor. This has allowed the study of the evolution of the nuclear wave function in several isotopic chains~\cite{datorresg-lanspaX}.

In recent years special efforts have been made to develop the use of the TF technique to study the $g$ factors of radioactive nuclei. In those works two low-spin population mechanism of radioactive nuclei can be mentioned: first, Coulomb excitation reactions with radioactive beams of long half life (they can be considered stable) and enough intensity. Second, the alpha-transfer reactions using a stable beam to create radioactive species that, otherwise, cannot be produced with enough intensity in the current radioactive beam facilities. In this contribution a short review of the experimental techniques utilized to populate low spin states in the measurements of magnetic moments in radioactive nuclei, using the TF technique, will be done. Some comments about perspectives and challenges to extend the experimental techniques to regions far away from the stability line will be mentioned at the end of the contribution.

\section{Population Mechanism of Low Spin States in Nuclei}
A complete review of the implementation of the TF technique to measure magnetic moments can be found in Refs~\cite{0954-3899-34-9-R01,kumTF2015}. Table~\ref{tab:table1} shows a list of some representative $g$-factor values measured with both techniques, the most interesting cases are those in which a pair of isotopes have been measured with a change in the sign for the $g$ factor. This is a clear signature of the influence of neutrons in the structure of the nucleus, see for example the isotopes $^{38,40}$S and $^{82,90}$Sr in Table~\ref{tab:table1}. This change of sign is pivotal to place stringent constraints on theoretical calculations of nuclear structure models~\cite{datorresg-lanspaX,0954-3899-34-9-R01,Speidel2002}.  
\begin{table}
  \centering
      {\small
\begin{tabular}{lcc|lcc}
\hline
\multicolumn{3}{c|}{Radioactive Beam} & \multicolumn{3}{c}{Alpha transfer}\\
Isotope & [ref] & $g(2^+_1)$ & Isotope & [ref] & $g(2^+_1)$ \\
\hline
$^{132}$Te  & \cite{BenczerKoller2008241} & $+0.28(15)$             & $^{68}$Ge & \cite{PhysRevC.71.044316}    & $+0.55(14)$\\
$^{38,40}$S & \cite{PhysRevC.74.054307}   & $+0.13(5)$, $-0.01(6)$  & $^{44,52}$Ti & \cite{Speidel2006219,Schielke2003153}        & $+0.83(19),+0.52(15)$\\       
$^{72}$Zn   & \cite{PhysRevC.85.034334}   & $+0.18(17)$             & $^{100}$Pd& \cite{PhysRevC.84.044327}   & $+0.30(14)$ \\
$^{76}$Kr   & \cite{Kumbartzki2004213}    & $+0.37(11)$             & $^{88}$Zr & \cite{PhysRevC.85.044322}    & $+0.30(11)$ \\          
$^{126}$Sn  & \cite{PhysRevC.86.034319}   & $-0.25(21)$             & $^{82,90}$Sr  &\cite{PhysRevC.89.064305}  & $+0.44(19)$, $-0.12(11)$\\
\hline
\end{tabular}
}
\caption[MagneticMoments]{\itshape $g$-factor values for the $2_1^+$ states for some representative radioactive isotopes, the values were obtained using the Transient Field technique in inverse kinematics. The experiments make use of Coulomb excitation with radioactive beams, left, or $\alpha$-transfer reaction, right, for the population of low spin states. }
\label{tab:table1}
\end{table}

\begin{figure}[ht!]
  \begin{center}
    \includegraphics[width=105mm]{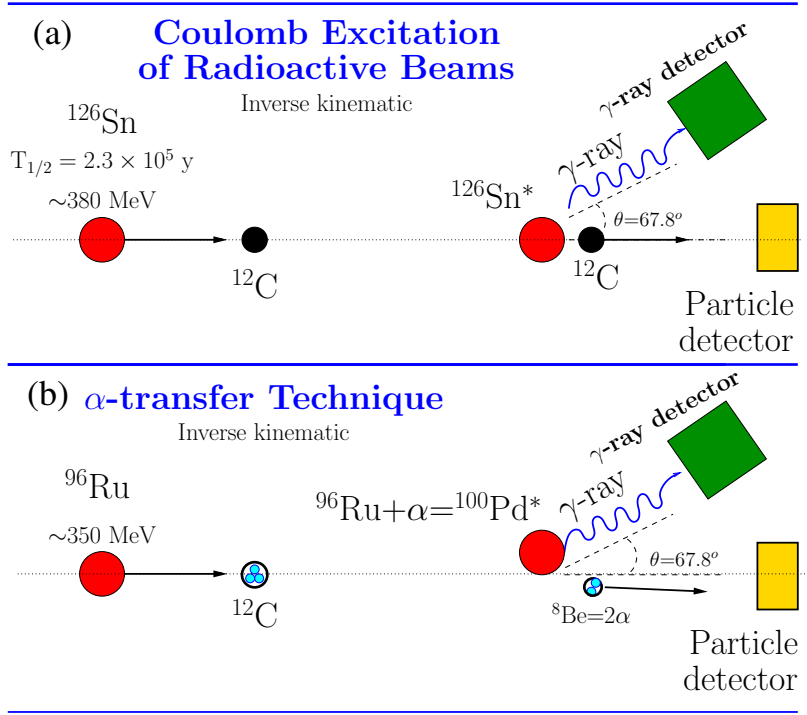}
   \end{center}
  \caption{\itshape Mechanisms to populate low-spin states of radioactive nuclei using the Transient Field technique in inverse kinematics. Upper figure {\bf (a)}, a coulomb excitation reaction populates low-spin states on the radioactive $^{126}$Sn beam nuclei, the states are produced with a high degree of nuclear spin alignment, a necessary condition to use the TF technique~\cite{PhysRevC.86.034319}. Lower figure {\bf (b)}, a stable beam of $^{96}$Ru pick up an $\alpha$ particle from $^{12}$C on the target; the populated states have a low degree of spin alignment, but it is still enough for the use of the TF technique~\cite{PhysRevC.84.044327}. }\label{montaje}
\label{fig:fig1}
\end{figure}

In Coulomb excitation reactions a radioactive beam is excited using a carbon layer to populate the low spin states of the beam. In even-even nuclei the $2^+$, and in some cases $4^+$ states, are populated with a high degree of spin alignment and a good yield production for the use of the TF technique. In the case of $\alpha$-transfer reactions a stable beam picks-up an $\alpha$ particle from a carbon nucleus in the target, a radioactive nucleus is created with certain degree of alignment to apply the TF technique. Figure~\ref{fig:fig1} presents a graphical representation of these two population mechanisms.

Three important points should be addressed if the use of the two reactions want to be extended to nuclei away from the stability line. They are the half life of nuclei under study, the spin alignment of the states and the TF parametrizations.

\subsection{Half life of the radioactive nuclei}
The production of radioactive nuclei implies the production of decaying sub-products, that may appear as contaminants for the purposes of the experiment. The treatment of such contaminants depends completely on the half life of the parent nuclei. For nuclei with half life longer than days no special requirements are needed to get ride out of the contaminants, they can be stopped in the back part of the target, the $\gamma$-decay from the sub-products will have a low effect on the final result, and in most of the cases the $\gamma$-ray decay does not interfere with the $\gamma$-ray line under study, as is the case of $^{126}$Sn~\cite{PhysRevC.86.034319} and $^{100}$Pd~\cite{PhysRevC.84.044327}.

A different approach has to be taken for the decay of short half-life states, of the order of hours, and high intensity, or around 10$^7$ to 10$^8$ particles per second, the use of a moving tape to take out the decaying sub-products from the reaction chamber is a solution that can be implemented, see for example the $^{76}$Kr experiment~\cite{Kumbartzki2004213}. 

\subsection{Spin alignment}

Coulomb excitation reactions produce the population of primarily $2^+$ states with a large spin alignment, this is the main reason why coulex have been the favorite reaction for TF experiments. The difficulty to populated  spin states with $J \ge 2$ is a limitation when coulomb excitation reactions are utilized.

Alpha-transfer reactions make use of a different population mechanism, and medium spin states with $J\ge 2$ are also populated with enough intensity to be studied. Unfortunately, a low spin alignment is produced in the states, resulting in large errors for the $g$ factors. The fact that states with low $J$ are fed by upper states implies that corrections are needed in order to take into account the precession from upper states. Two interesting points are that $\alpha$-transfer reactions produce better alignment in upper states, and that states with $J>6$ seem not being populated in the reactions. 

\subsection{Transient Field Parametrization}

The lack of a model to obtain precise values for the intense magnetic field, produced by the spin-orbit interaction utilized by the TF technique (of the order of kilo Tesla), is one of the most important challenges to extend the TF technique to nuclei away from the line of stability~\cite{kumTF2015}.

\section{Conclusions and perspectives}

The use of the Transient Field technique to measure magnetic moments in nuclear states with short lifetime has permitted to contrast several nuclear structure models along the chart of nuclides in the last decades. To date, the main studies have been concentrated in the measuring of $g$-factor values of $2^+$ states in stable even-even nuclei. Studies in forthcoming years will concentrated their efforts, primarily, in the characterization of low spin states of radioactive nuclei using coulomb excitation reactions. Such studies will take advantages of the new radioactive beam facilities and their capacity to develop and deliver radioactive beams, with half life of the order of hours and more, with enough intensity and beam energy to take advantage of the TF technique. The use of radioactive beams with short half life is one of the challenges to be experimentally addressed.

A second front is the study of spin states with $J\ge 4$, that could be populated making use of $\alpha$-transfer reactions. The study of the reaction mechanism in the $\alpha$-transfer, and the population mechanism of the states of the products of the reaction, are some of the challenges that have to be studied both theoretically and experimentally. It is important also to point out that a limited number of $\alpha$-transfer experiments using stable beams are worth to be performed, and most of the cases have been already done. The possibility to use the $\alpha$-transfer mechanism with radioactive nuclei open a very interesting challenge, that should be explore in the near future.

Finally, the study of the TF technique parametrization itself is probably the main obstacle that should be focus on the forthcoming years. Such studies should take advantage of the powerful theoretical tools developed during the last decades, to replace the utilized parametrizations with a theory from first principles.

\subsection*{Acknowledgements}

This work has been supported in part by Colciencias under contract 110165842984.

\bibliographystyle{unsrt}
\bibliography{DiegoTorres}
\end{document}